\documentclass[11pt]{article}

\usepackage[margin=1in]{geometry}
\usepackage{amsmath}
\usepackage{amssymb}
\usepackage{graphicx}
\usepackage{booktabs}
\usepackage{array}
\usepackage{tabularx}
\usepackage{microtype}
\usepackage{authblk}
\usepackage{xcolor}
\usepackage[flushmargin,hang]{footmisc}
\usepackage[hidelinks,colorlinks=true,linkcolor=blue,citecolor=blue,urlcolor=blue]{hyperref}
\usepackage{xurl}
\usepackage{caption}

\title{\textbf{Repurposing Image Diffusion Models for Adversarial Synthetic Structured Data: \\ A Case Study of Ground Truth Drift}%
\thanks{An earlier version of this research was presented by Adam Arthur, ``Repurposing Image Diffusion Architectures for Tabular Synthesis,'' the Rochester Institute of Technology Undergraduate Research Symposium, Rochester, NY, July 31, 2025.}}

\author[1]{Adam Arthur}
\author[1]{Christopher Schwartz}
\affil[1]{Rochester Institute of Technology \\ \texttt{aia5138@rit.edu} \quad \texttt{ccsics@rit.edu}}

\begin{document}
\maketitle

\begin{abstract}
\noindent
Public image diffusion models are now powerful enough that an attacker without the resources to train a tabular-specific generator may repurpose one off the shelf. This study tests that possibility directly. An unmodified Stable Diffusion U-Net is applied to the UCI Adult Income dataset by reshaping each row into a small single-channel pseudo-image. The architecture's inductive bias toward spatial locality makes feature placement a design variable, and several layouts are tested. However, this is only the beginning of the story, as this paper also draws two philosophical distinctions. One separates statistical from perceptual realism: whether synthetic content holds up to a machine's correlation audits or a human's sensory inspection. The other introduces synthetic evidence as a category alongside synthetic media: AI-generated material whose consumer is a machine in a closed evidentiary pipeline rather than a person in an open information system. An attacker succeeds with synthetic evidence by thinking like the machine that will receive it. And the more the attacker succeeds, the more they can induce ground truth drift: the silent reclassification of AI-generated outputs as authentic when reused in pipelines that do not interrogate their provenance.
\end{abstract}

\pagebreak


\section{Introduction}
\label{sec:Introduction}

Synthetic content is most often discussed as a problem of human perception: deepfaked faces, fabricated voices, doctored news footage. This paper takes up a different and less examined case, where the audience for the deception is not a person but a machine, and the artifact in question is not a video or an image but a row in a database. 

The motivations for this study are twofold. In the immediate term, the threat surface around such content is shifting. A few years ago, fabricating convincing tabular records such as synthetic patient histories and synthetic financial transactions required either dedicated tabular generators such as CTGAN or TabDDPM, or hand-crafted statistical methods. Today, it is now possible for an attacker who never trains a model from scratch to pull a high-capacity perceptual diffuser off the shelf and make it do the job. 

The threat surface is ultimately more philosophical than it is technical: a phenomenon we call \emph{ground truth drift}. The term names a category that is adjacent to but distinct from \emph{model collapse}, which describes the iterative degradation that occurs when generative models are repeatedly trained on the outputs of their predecessors, with the tails of the distribution thinning out over generations.~\cite{shumailov2024} Ground truth drift, by contrast, names two phenomena. The short-term phenomenon is what happens at any single point of reuse: a synthetic artifact enters a downstream training pipeline, the pipeline does not know it is synthetic, and the categorical shift from ``generated'' to ``authentic'' happens silently, without recursion and without distributional change. The long-term phenomenon is the cumulative deleterious effect such incidents have for both people and machines: people lose access to a reliable evidential substrate for their decisions, and machines are trained on a slowly accumulating sediment of artifacts that no one ever vouched for.

The synthetic artifact in question does not need to be malicious to do damage; mislabeling, misinterpretation of labels, undisclosed provenance, or simple negligence will do. Take for example a 2021 dataset of synthetic medical images produced by Bradley Segal et al,\cite{segal2021} and distributed through the Medigan python library.\cite{medigan_repo} The dataset contains a substantial fraction of images carrying the label ``no finding.'' This is per the NIH labeling schema, where ``no finding'' indicates the absence of 14 specific pathological conditions.\cite{badr2022} The slippery part is that despite explicitly marking the presence or absence of pathologies, the NIH treats ``no finding'' as a technical label, not a clinical one. Whether the depicted chest is healthy (or ``healthy'' in the case of synthetic X-rays), or whether the image is even a viable basis for diagnosis (or diagnosis's simulated equivalent), is not what the label captures. A study with severe artefacts, a poor-quality mobile chest X-ray, or non-target pathology can all legitimately receive the ``no finding'' label.~\cite{segal2026personal} A downstream researcher who pulls these images into a training set and treats ``no finding'' colloquially as ``healthy'' rather than as the technical category it is, is now training a model under a label whose meaning has shifted across the consumption boundary.~\cite{segal2026personal} 

The question this paper takes up is what happens when drift is deliberate, i.e., when an adversary, rather than a well-intentioned researcher, is the source of the contamination. The case study presented in this paper -- an unmodified Stable Diffusion U-Net producing tabular data that passes standard statistical audits while harbouring systematic logical violations -- is one concrete instance of how cheaply such adversarial drift can now be manufactured.

\subsection{Overview}

While the motivations for this study are temporal, the design is fundamentally conceptual. At present, synthetic content is taxonomized in a confusing manner: it is possible to generate ``synthetic data,'' which can be either ``structured'' or ``unstructured,'' which is to say, tabular or audiovisual in nature. Unstructured synthetic data is also referred to as ``synthetic media,'' but this term has a dual scientific and forensics connotation. When the latter, it is also known more popularly as ``deepfakes.'' We find synthetic media too unhelpful to capture the problem of ground truth drift, and so we introduce a new term: ``synthetic evidence.'' 

From a cybersecurity perspective, both synthetic media and evidence share the intent to persuade and deceive, but they differ in their audiences and ecosystems. Synthetic media targets people and propagates in porous information communication systems such as online social networks, whereas synthetic evidence targets AI models and propagates in closed evidentiary pipelines such as those found in science. From a philosophical perspective, synthetic media and evidence diverge because of a difference between \emph{statistical} versus \emph{perceptual realism}. Statistical realism concerns whether synthetic content holds up to a machine's correlation-and-attribute audits, whereas perceptual realism concerns whether it holds up to a human's sensory inspection.

This distinction between two kinds of realism also motivates the concept of \emph{architectural convertibility} that is at the core of our studied attack. Convertibility occurs when a model built for one kind of realism (e.g., perceptual) is successfully repurposed to produce another (e.g., statistical) without modification (i.e., fine-tuning). The more convertible a model -- especially an off-the-shelf model -- the cheaper the cost of mounting a drift attack becomes.

The threat model assumed throughout this paper is a low-resource but technically capable adversary. They lack the data, compute, or expertise to train a tabular-specific model. What they have is access to public image diffusion models, knowledge of how a target pipeline ingests tabular data, and a willingness to experiment with input reshaping. Their goal is not perfect record forgery. Rather, it is volume: flooding a downstream system, such as a fraud-detection classifier, a population-statistics dashboard, or a benchmark dataset, with synthetic rows that pass the kinds of automated audits commonly applied to tabular data. Whether such an adversary can succeed using only a stock image diffuser, with no architectural modification, is the question this paper takes one concrete step toward answering.

The technical move that makes the rest of this work possible is a reshape. Stable Diffusion's U-Net expects an image-shaped tensor: channels by height by width. A tabular row is a flat feature vector. By scaling the numerical columns, one-hot encoding the categorical ones, and laying the resulting vector onto a small fixed grid as a single-channel tensor, each row becomes a \emph{pseudo-image}: a 10 × 11 picture whose pixels happen to encode a census record. The model itself is left completely unchanged. It does not know the input is tabular; it learns the distribution of these small pictures the same way it would learn any other image distribution. At inference, a new pseudo-image is sampled from noise and the cells are read back in a fixed order to recover a synthetic row. 

The trick is to think like the machine. It sees numbers, not images, and it has a strong inductive bias toward spatial locality: convolutional layers act on local neighborhoods, which means features that need to constrain each other -- sex and relationship, education and occupation -- must sit near each other on the grid for the model to learn their relationship. The layout of the row on the grid is therefore a design variable, not a convenience. Different layouts produce meaningfully different outputs from the same network.

The experiments that follow probe how far this trick goes. Across three feature layouts -- arbitrary, correlation-clustered, and manually grouped by semantic relatedness -- the same unmodified U-Net produces synthetic data ranging from clearly broken to surface-plausible. Statistical fidelity, measured by SDMetrics, sits in the mid-80s percentile across all configurations. Logical consistency tells a different story. Row-level logic errors drop from roughly 70 percent under arbitrary layout to about 12 percent under manual grouping, and then plateau. Further layout adjustment does not close the remaining gap. The headline finding is that an unmodified vision U-Net can convincingly mimic the surface statistics of structured data while reliably producing rows that fail basic semantic checks, e.g., negative capital gains. From a security standpoint, this is exactly the gap an attacker would want: the data passes audits that look at columns, and fails audits that look at rows. From an architectural standpoint, it delineates how far perceptual-model reuse can be pushed without modification, and where the wall is.

\subsection{Procedure}

Section~\ref{sec:Philo-Background} develops the philosophical framing, including some thoughts on categories of synthetic content/artefacts, the contrast between persuading people and persuading machines, and a fuller treatment of ground truth drift than the introduction has space for. Section~\ref{sec:Attack} conducts the attack: surveying the architectural landscape that makes it possible, identifying the specific gap this paper exploits, and describing the experimental setup in detail -- dataset, tensor mapping, and the three layout strategies. Section~\ref{sec:Data-Analysis} presents the results across statistical fidelity, semantic consistency, and downstream classifier utility. Section~\ref{sec:Discussion} interprets the findings, with the Francesca Gino case serving as the concrete real-world anchor for what an adversarial instantiation of ground truth drift would look like in practice.

\section{The Philosophical Attack Surface}
\label{sec:Philo-Background}

\begin{figure}[!ht]
\centering
\includegraphics[width=0.5\linewidth]{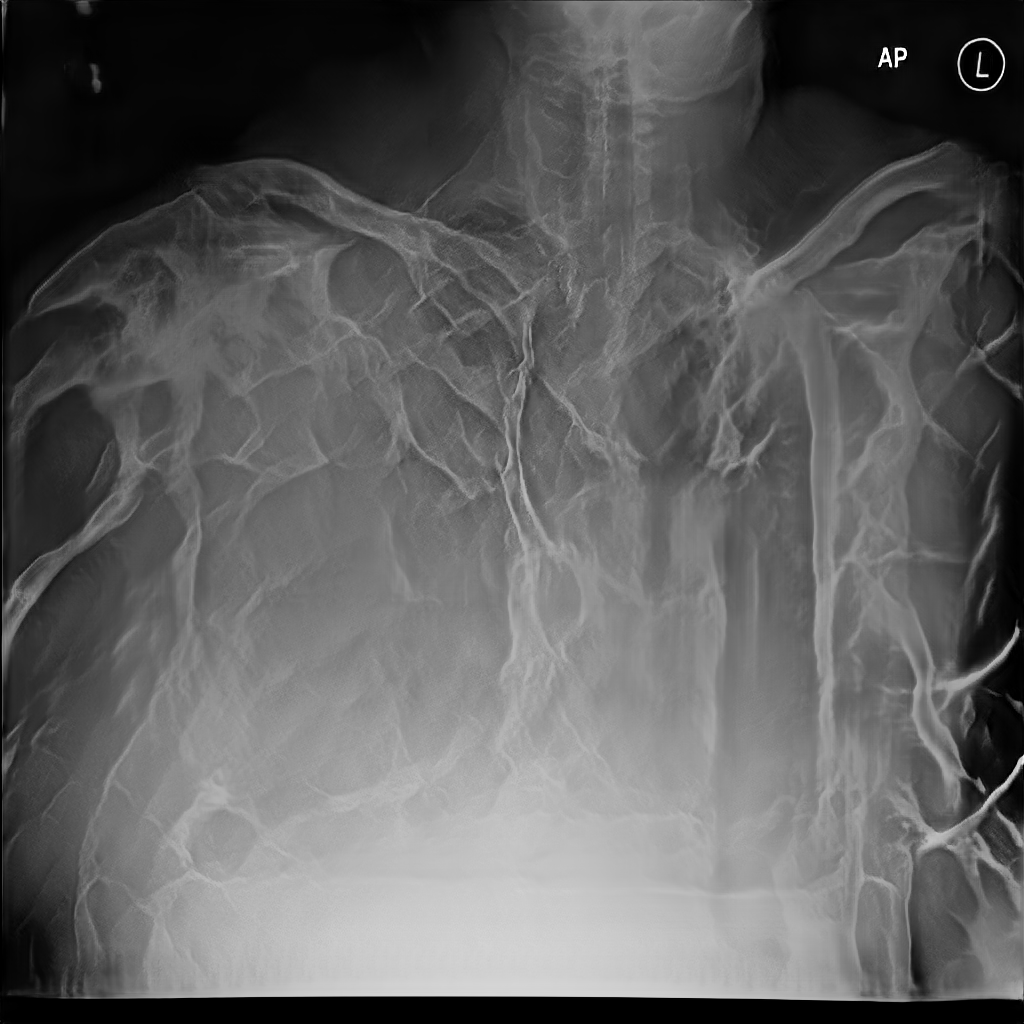}
\caption{A synthetic chest X-ray from the 2021 Segal et al. dataset~\cite{segal2021}, labeled ``no finding'' within the NIH CXR labeling schema. The dendritic filaments fanning across both hemithoraces are an instance of generator collapse: the GAN produced an output where lung fields and a cardiac silhouette would be expected.}
\label{fig:ground-truth-drift-1}
\end{figure}

\begin{figure}[!ht]
\centering
\includegraphics[width=0.5\linewidth]{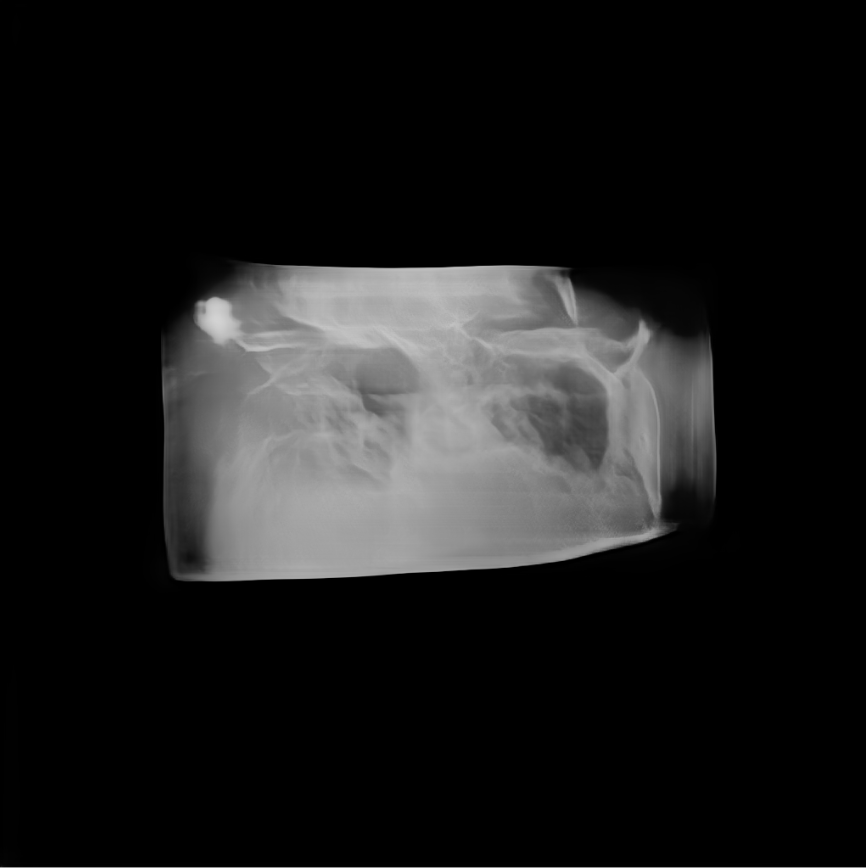}
\caption{A second synthetic chest X-ray from the same dataset, also labeled ``no finding.'' The image reproduces the visual idiom of a poor-quality mobile chest radiograph, where ``no finding'' is consistent with the source schema but does not indicate a healthy chest.~\cite{segal2026personal}}
\label{fig:ground-truth-drift-2}
\end{figure}

While synthetic data and media as \emph{practices} predate generative AI by several decades, the actual usage of these specific terms has varied in response to technological advancements. In contemporary usage, ``deepfake,'' ``synthetic media,'' and ``synthetic data'' are all strongly associated with AI-generated content, which makes it hard to differentiate between them. The Royal Society~\cite{royalsociety2024} notes that no widely accepted definition currently exists, tentatively describing synthetic data as any information generated using a purpose-built model or algorithm to solve a data science task. For example, NIST SP 800-188~\cite{nist2023} defines synthetic data as artificial data that share some statistical characteristics with seed data -- a definition that covers structured, non-audiovisual content but does not account for synthetic media, even though the latter would also fit the prima facie reading of that definition. In contrast, an executive order from the US President groups AI-generated images, audio, and video under the catch-all banner of ``synthetic content''~\cite{eo14110}. The European Data Protection Supervisor's TechSonar entry~\cite{edps2022} similarly groups image, audio, and tabular synthesis under a single ``synthetic data'' header, while emphasizing privacy use cases over evidentiary ones.

This definitional imprecision can be traced back to how each of these terms first entered the technical and public discourse. The earliest of the three, synthetic data, seems to have entered formal use in 1987, when Donald Rubin proposed creating artificial tabular records that mimic a real dataset's statistical properties to protect individual privacy. During this time, synthetic data was understood to be artificially created structured or tabular data with the intent to model an original dataset's statistical properties. The defining features required that it was not merely arbitrary fake data, but rather data generated by a model fitted to the real data. There was no inclusion of AI generation techniques in its methods, and the outputs were specifically related to datasets and did not include audiovisual content. Synthetic media was not a known phrase at the time but the concept of creating ``fake media'' like CGI mirrored the idea of deepfaking media. As technology began to modernize, common terminology for unstructured synthetic data was mostly restricted to ``synthetic images,'' ``image synthesis,'' and ``GAN-generated faces/images.'' There was no formal umbrella term concerning synthetic forms of media, and synthetic data still referred to structured datasets.

This all began to change when the infamous case of a Reddit user named \emph{deepfakes} posted AI-swapped adult content. The term deepfake immediately caught public attention and began to develop a negative connotation. Synthetic media was not popularized at that time, but the following year there was a need for a term to categorize artificial media without the negative connotations of ``deepfake.'' Consequently, synthetic media began to be widely adopted in formal contexts, including academic literature and policy/regulations, to refer generally to forms of synthetic content intended for human consumption. It was easier to use synthetic media in formal contexts because it sounded broader, safer and not tied to deception like ``deepfake.'' In public discourse, any form of artificially generated fake content still was hastily marked as a deepfake especially when it pertained to human faces and voices.

A further wrinkle complicates the vocabulary still further: synthetic media \emph{is} synthetic data. Both consist of artificially generated values produced by a model fitted to real source material, and the only durable difference between them is one of structure. \emph{Structured} synthetic data inhabits a tabular form -- rows, columns, fixed types -- where each cell carries a discrete, named meaning. \emph{Unstructured} synthetic data inhabits the form of pixels, waveforms, or token sequences, where meaning is distributed across the whole signal rather than tied to any one position. Synthetic media is, by this measure, simply unstructured synthetic data.

It would further follow that not every piece of synthetic media is necessarily a deepfake. Generative models routinely produce unstructured outputs whose intended audience is another machine or a closed scientific workflow, never a credulous human viewer. Synthetic chest X-rays can augment a medical classifier's training set; synthetic LiDAR scans can fill out the long tail of an autonomous-driving simulator; synthetic spectrograms can balance an audio-classification dataset; synthetic satellite imagery can stress-test a remote-sensing pipeline against rare weather conditions. None of these artifacts has a deceptive aim against a person, and treating them as deepfakes flattens a distinction that matters both technically and ethically. A deepfake, by contrast, is unstructured synthetic data specifically routed toward human consumption with the intent to mislead.

That said, the forensics profession evinces ambivalence about the term deepfake. Their preference is for precision, such as ``synthetic faces'' or ``facial forgeries'' (when what is being represented is a human face) or ``synthetic forgeries'' (as a broader term). Even so, forensics still regularly defaults to synthetic media. This causes the term to carry the implication of a distinct kind of artifact, one that carries forensic weight. And so, synthetic media currently is the vocabulary for \emph{both} investigators, journalists, and policy-makers asking who made a piece of content and to what end, and computer scientists asking how the content was generated. The distinction is important, but it is one of \emph{use}, not of \emph{kind}. It is also, in our view, still insufficient, which is why we introduce the further distinction between statistical and perceptual realism.

\subsection{Persuading People versus Machines}
\label{sec:people-vs-machines}

Synthetic content can be understood, à la Rubin, as AI-generated material that seeks to mimic real material's statistical properties. The goal behind doing so may not only be to preserve privacy, as Rubin suggested. For example, modern AI models often require a certain sheer amount of data to train on, while real data on a given topic or phenomenon may be sparse. Synthetic data can thus step into the gap and numerically supplement the real data.

Of course, synthetic content can also be used for the purposes of deception, but properly reckoning with this requires philosophical nuance. Anthropocentrism creeps in here, as whenever we think about deception, we immediately think of the deceived as being people. Yet, AI models themselves may be the target of deception. This is the essence of the distinction drawn in this study between perceptual and statistical realism: all synthetic content attempts to appear real, but the way it may appear so to a human mind is not necessarily the same way it would to an AI model.

Consider the seemingly straightforward classification task of identifying cats versus dogs. A human mind is likely to distinguish between these animals based on their biological features, while an AI model could very well identify them based on environmental cues. When thought of as \emph{pets} and not as \emph{wild animals}, cats will typically be found indoors, whereas dogs are frequently outdoors.\footnote{This is known as the Clever Hans Effect~\cite{cleverhans_wiki}. Here, rather than treating it as faulty or anomalous, we treat it as epistemologically meaningful. To a human mind, identifying cats versus dogs based on environmental cues may not be how we would identify these pets, but that does not \emph{necessarily} make it incorrect. If a classification task genuinely requires that it be performed in the way human minds would perform it, then it would be faulty; but if a classification task could reliably achieve as good or better results differently, then it could be justified.} The lesson here for the aspirant deceiver is that an AI model will latch onto a statistical feature rather than a perceptual one,\footnote{Of course, one can argue that both human and machine perception rely on probability, they just prioritize different priors. In the case of this example, biological ones for human minds, environmental ones for AI models.} so where people must be deceived by apparent \emph{sensory} high fidelity -- or at the very least, plausibility -- an AI model must be deceived by apparent \emph{mathematical} conformity.\footnote{Of course, ``short circuiting'' is also a possibility, as in the case of adversarial examples.}

Which kind of deception matters depends on the audience and ecosystem of the synthetic content. Content destined for a closed, evidence-based pipeline -- a scientific experiment, a fraud-detection model, an automated audit -- will be consumed by machines, and statistical conformity is what determines whether it slips through. Content destined for an open, porous media system -- a social network feed, a courtroom exhibit, a news segment -- will be consumed by people, and sensory plausibility is what determines whether it is believed. Synthetic data, in the narrow tabular sense, should therefore be judged by the standards of statistical realism, while synthetic media is associated with perceptual realism. 

Statistical realism is focused specifically on making the output data resemble the original in its correlation and attributes; this makes it perfect for AI training data. The defining factor lies in its ability to artificially simulate real data for models and analysts, disregarding in what ways a human could perceive them. Negligent of the characteristics of the underlying statistics, perceptual realism is instead focused solely on how the output would appear to a human. For such a purpose, artificially generated audiovisual content is typically categorized with perceptual realism. 

\subsection{Ground Truth Drift}
\label{sec:ground-truth-drift}

The chest X-ray case introduced earlier illustrates a deeper structural problem: the boundary between synthetic media and synthetic evidence is not only blurry at the point of generation but also at the point of \emph{re-use}. A piece of content created as synthetic media for one purpose can quietly become synthetic evidence for another. A synthetic chest X-ray, generated to augment a perceptual training set or to anonymize a patient record, is consumed by a downstream classifier as a labeled instance in that classifier's ground truth. The image was never published in any conventional sense -- it never circulated through any of the channels that ``media'' implies -- but it now sits in the training pipeline as evidence, contributing to the model's notion of what a real X-ray ``should'' look like.

This is what we mean by ground truth drift. The same piece of content can be either media or evidence, depending on whether its maker intends to use it for scientific or persuasive purposes, and on whether the environment it ends up in is closed or open. Drift becomes a problem when the maker's intent and the environment's reception fall out of alignment: when content produced to persuade ends up training a model, when content produced to train a model ends up persuading a person, or when content produced as a perceptual exercise ends up serving as evidence in a closed scientific pipeline. The classification of synthetic content cannot be settled at the moment of generation; it has to be re-evaluated at every downstream consumer.

The chest X-ray case illustrates drift by accident: there was no adversary, only the diffusion of well-intentioned synthetic material into pipelines whose interpretive conventions widened the schema's specificity beyond what the original labels supported. The present study illustrates the corresponding case of drift by design. Where the X-ray dataset became drifted evidence through neglect of label semantics across pipeline boundaries, the tabular rows produced by an unmodified Stable Diffusion U-Net are deliberately constructed to look like authentic data under the audits commonly applied to it. The technical achievement is modest -- a reshape, a layout choice, an off-the-shelf model -- but the implication is not. If a low-resource adversary can manufacture statistically conformant synthetic evidence at volume, the boundary between media and evidence has collapsed not just at the point of re-use but at the point of generation. For the present study, the implication is narrow but pointed: a tabular row generated by a perceptual model is not safely categorizable as either synthetic media or synthetic evidence in advance.


\section{Conducting a Synthetic Structured Data Attack}
\label{sec:Attack}

This section presents the experimental component of the paper. Where Section~\ref{sec:Philo-Background} argued that synthetic evidence and synthetic media are different categories of artefact deceiving different audiences, the work that follows shows what one such attack actually looks like in practice: an adversary equipped with public infrastructure -- an off-the-shelf image diffusion model, a consumer GPU, a tabular benchmark dataset -- successfully producing structured records that pass the column-level audits commonly applied to such data. The attack is interesting not because it is technically sophisticated (it is not) but because of how cheap it has become. We first survey the architectural landscape that motivates the attack and identify where existing literature has and has not anticipated what we are doing (Section~\ref{sec:Tech-Background}). We then describe the experimental setup in detail (Section~\ref{sec:Methods}).

\subsection{Generator Families and Architectures}
\label{sec:Tech-Background}

Among the architectures most often used in synthetic media, the U-Net is the most relevant to the present study. A U-Net is a convolutional neural network with an encoder and a decoder linked by skip connections. The encoder compresses an input into higher-level feature maps, and the decoder upsamples and refines those features back toward full resolution using the skipped outputs. Inputs and intermediate representations are tensors, that is, multidimensional arrays; an image is typically height by width by channels. Convolutions act on local neighborhoods within these tensors, so the architecture has a built-in bias toward spatial locality and smooth structure. That bias aligns with synthetic media, where perceptual realism in images benefits from spatial correlations. Synthetic data in tabular form does not have a canonical spatial layout: column order is often arbitrary, and many dependencies are long-range or combinatorial rather than local. These contrasts motivate the question of whether an image-oriented architecture can capture the statistical structure expected of synthetic data.

The generator families behind synthetic data and synthetic media have begun to overlap as both have advanced. The two domains converged when GANs, diffusion models, and transformer architectures, originally developed for image synthesis, were extended to structured data generation~\cite{goodfellow2014, ho2020, vaswani2017}.Researchers have applied these models to medical imaging including X-rays and MRIs, tabular data such as CTGAN and TabularGAN, and time series data including sensor streams~\cite{shin2018, xu2019, esteban2017}. The same core architectures are now used to generate both structured and unstructured outputs. As a result, companies offering AI generation tools for multiple output types have begun marketing themselves as synthetic data platforms.To align with commercial use cases, such firms often categorize synthetic tabular data, medical images, videos, and speech collectively as synthetic data~\cite{gartner2022}. This broader industrial usage preceded uptake in academic literature.

Within structured domains specifically, contemporary synthetic data pipelines rely on essentially the same generative engines that first revolutionized image synthesis. GAN-based frameworks remain dominant. CTGAN and related variants train an adversarial pair in which a generator proposes full rows of data while a discriminator flags implausible combinations, iteratively refining distributions until synthetic tables are statistically interchangeable with the originals~\cite{xu2019}. This approach has gained popularity in finance research, where similar architectures are used to simulate credit card streams, stock price trajectories, and other time series with realistic properties. Diffusion models, well known for generating high-fidelity images, have been repurposed as TabDDPM, a tabular denoising process that transforms Gaussian noise into coherent records through a sequence of refined predictions~\cite{kotelnikov2023, villaizan2024}. Applying a model originally intended to reconstruct progressively higher-quality image stages toward the task of generating spreadsheet rows illustrates how far these core architectures have generalized.

Recent efforts to repurpose generator models across tasks typically rely on fine-tuning rather than retraining from scratch, and most of this work remains within the image domain. Work with the StyleGAN family illustrates this pattern, with models retrained for new visual tasks, including medical imaging such as CT, MRI, and mammography, while the architecture is reused and weights are adapted to the target data. Across these cases, the architecture is retained but the weights are retrained on the target domain. We are not aware of prior work testing exactly this attack pattern. Domain adaptation studies generally remain within a single modality, typically covering image-to-image translations such as MRI to CT using CycleGAN, or stylistic transformations like photorealistic to cartoon faces. There is little evidence of a vision model trained for facial synthesis being applied in unmodified form to structured data. In the text domain, large language models trained on general corpora have been adapted to clinical notes and synthetic patient data, though this remains within language. One recent exception is CTSyn~\cite{lin2024}, which uses a diffusion autoencoder trained on tabular datasets to synthesize structured data across use cases. This represents an effort to develop a foundation model for tabular data, analogous to how GPT and Stable Diffusion function for text and image generation respectively, though CTSyn is trained directly on tables rather than transferred from another domain.

Diffusion U-Nets, which are widely used in image generation, are currently being redesigned for tabular contexts through multilayer perceptrons or custom noise conditioning for categorical variables. However, no studies have tested whether a vision-based U-Net can generate structured data without architectural modification. Prior studies report that TabDDPM outperforms CTGAN, but the performance of a raw U-Net has not yet been established. Whether such a model can learn statistical distributions without structural adaptation remains an open question. This paper takes that question head-on by testing whether an unmodified vision-based U-Net, originally built for perceptual tasks like image generation, can be directly applied to generate structured tabular data with statistical realism.

\subsection{Methods}
\label{sec:Methods}

The experiment is a single architectural sleight-of-hand followed by a search for its ceiling. An unmodified Stable Diffusion v1.5 U-Net is trained on the UCI Adult Income dataset, with each tabular row reshaped into a 10 × 11 × 1 single-channel tensor -- a small pseudo-image whose pixels encode the row's features. The model itself is never changed; only the spatial arrangement of features within the grid is varied, across three layouts (arbitrary baseline, correlation-clustered, and manually grouped by semantic relatedness). Each configuration is then evaluated on the same battery of metrics covering statistical fidelity, semantic consistency, downstream classifier utility, and disclosure risk. The subsections that follow describe the dataset and training setup, the tensor mapping that makes a row image-shaped, the three layout strategies, and the evaluation criteria in turn.

\subsubsection{Dataset and Training Setup}
\label{sec:Dataset-and-Training}

To test whether an architecture built for perceptual realism could generate statistically realistic structured data, the unmodified U-Net backbone from Stable Diffusion was applied to the UCI Adult-Income dataset. Since the model was designed for image data, each tabular row was reshaped into a format it could process without changing the architecture itself. The experiment also compared multiple feature layout strategies such as baseline, clustered, and manual to assess how data formatting influences model performance and whether the same architectural limitations appear across all configurations.

The UCI Adult Income dataset was used for training and evaluation, a census style table with about 48{,}000 records. Each row includes age, education, occupation, marital status, race, sex, hours worked per week, native country, and the income label. It is a standard tabular machine learning benchmark and an appropriate reference when comparing an image model to table native generators. Its mix of numerical variables (age, capital gain or loss) and categorical variables (relationship, education) allows evaluation of distributional fidelity and semantic consistency. The model was the Stable Diffusion v1.5 U-Net from the Diffusers library and the architecture was left unchanged. Training used the library diffusion loop with AdamW, the default learning rate and batch size, and the provided noise schedule for 50 epochs on a local RTX 3060. Samples were generated every five epochs and produced 5{,}000 synthetic rows per run. No hyperparameters were tuned.

\subsubsection{Input Formatting and Tensor Mapping}
\label{sec:Tensor}

For the purposes of this experiment, the model did not learn from the data in its typical format. Categorical columns were one-hot encoded and numerical columns were scaled, then concatenated into a single feature vector. Typical synthetic data models are designed to take such a vector, while image models are designed for tensor input. Both approaches convert data to numbers, but by different means: vectors and tensors. To bridge this, each row was mapped to a $10 \times 11 \times 1$ tensor by placing feature values on a fixed grid with a single channel, forming a small fake image that preserved the scaled information from the original table. Inputs were normalized to the range expected by the pipeline.

\subsubsection{Layout Design Strategies}
\label{sec:Layout}

Column placement within the input tensor affected the outputs based on how the training data was formatted. In an effort to maximize training effectiveness, three spatial layouts were defined that place certain features near one another to help the model learn their correlations and behaviors. The first attempted layout was a baseline with values ordered arbitrarily or for processing convenience, with no deliberate structure. Next, clustered layout was used. It grouped features by pairwise correlation computed on the training data and placed those groups as neighbors on the $10 \times 11$ grid to help the network learn local dependencies. Finally, manual layout was used that positioned semantically related fields next to each other based on dataset context, for example sex next to relationship status and education near occupation. Together with the baseline, these layouts defined the spatial arrangements tested in this study.

\subsubsection{Evaluation Criteria}
\label{sec:Eval-Criteria}

Each layout was evaluated on 5{,}000 generated rows. Statistical fidelity was measured with SDMetrics using column shape scores and pairwise correlations. Semantic consistency was checked with a rule set that flagged invalid numeric ranges and contradictory labels, including capital gain at or above zero, hours worked within a realistic range, sex and relationship consistency, and education within the valid one to sixteen scale. Downstream utility was tested with a train-on-synthetic, test-on-real income classifier, and F1 was reported for the greater than 50K class. Privacy risk was estimated with a disclosure score that quantifies similarity between synthetic and real records to assess potential memorization.

\section{Data Analysis}
\label{sec:Data-Analysis}

Each layout was evaluated on five thousand generated rows, and the results split cleanly along the statistical-versus-semantic axis introduced earlier. On surface metrics -- column shapes, pairwise correlations, overall SDMetrics scores -- every configuration cleared the 80 percent threshold, and the clustered layout reached 86.63 percent. On row-level logic, the picture changes: arbitrary placement produced inconsistencies in roughly 70 percent of rows, semantic grouping cut that to about 12 percent, and further layout adjustment did not close the remaining gap. The classifier evaluation tells a third story still, with strong performance on the majority income class and persistent failure on the minority class across every layout. The subsections below work through these results in turn: first the headline numbers, then the fidelity and semantic findings in detail, then the classifier and feature-distribution analysis that explains why one layout outperforms the others on downstream tasks.

\subsection{Quantitative Overview}
\label{sec:Quantitative}

Across the three experiments, each feature layout produced noticeably different outcomes directly related to how the data was formatted. The clustered grid achieved the highest statistical fidelity with an SDMetrics score of 86.63\%, whereas the baseline came in lowest at 81.89\%. None of the layouts, however, had been shown to consistently connect character traits across columns to a standard degree. From a human perspective, it is evident that in a select number of rows, the data contains inconsistencies and logically impossible combinations. Through the evaluation of an automated logic check, the manual and clustered layouts held error rates of 11.72\% and 14.17\% respectively, while the baseline, due to its lack of normalization, grew to 70.08\%. In the Train-on-Synthetic, Test-on-Real evaluation, every layout handled the majority income group ($\leq 50{,}000$) well, averaging an F1-score of about 0.87 across configurations; by contrast, all layouts severely struggled with the minority group ($> 50{,}000$), where F1-scores fell between 0.06 and 0.25. Privacy risk remained generally low as disclosure scores clustered between 0.61 and 0.65, indicating the diffusion process did not memorize the training data.

\begin{table*}[ht]
\centering
\caption{Key Metric Comparison Table}
\label{tab:key-metrics}
\small
\resizebox{\textwidth}{!}{%
\begin{tabular}{@{}lccc@{}}
\toprule
\textbf{Metric} & \textbf{Baseline (Random)} & \textbf{Clustered + Normalized} & \textbf{Manual Grouping (Final)} \\
\midrule
\multicolumn{4}{l}{\textit{Statistical Fidelity (SDMetrics)}} \\
\quad Column Shapes          & 83.62\% & \textbf{91.59\%} & 90.42\% \\
\quad Pairwise Correlations  & 80.34\% & \textbf{81.66\%} & 77.85\% \\
\quad Overall SDMetrics Score & 81.98\% & \textbf{86.63\%} & 84.14\% \\
\addlinespace
\multicolumn{4}{l}{\textit{Downstream Usability (TSTR)}} \\
\quad Overall Accuracy        & 77.0\%  & \textbf{78.5\%}  & 77.6\% \\
\quad F1-Score ($\leq 50$K)   & 0.869   & 0.87             & 0.87 \\
\quad F1-Score ($> 50$K)      & \textbf{0.056} & \textbf{0.25} & 0.16 \\
\bottomrule
\end{tabular}%
}
\end{table*}

\subsection{Statistical Fidelity Deep Dive}
\label{sec:Fidelity}

Further SDMetrics raw-fidelity tests showed all layouts exceeded 80\%, indicating respectable statistical fidelity. The model particularly excelled in column shape scores, demonstrating competent handling of value distributions. Manual and clustered layouts scored above 90\%, while the baseline trailed at 83.62\%. Pairwise correlations ranged from 78.9\% to 81.4\%, suggesting most feature relationships were preserved despite some moderate discrepancies. Despite these high scores, SDMetrics cannot assess logical or semantic consistency, so deeper flaws may remain undetected.

\subsection{Semantic Consistency Analysis}
\label{sec:Semantic}

Semantic consistency was assessed with a custom logic checker that scanned each synthetic row for six recurrent violations: invalid numeric ranges such as negative capital gains and hours worked above one hundred, and education levels outside the valid one-to-sixteen scale. In the baseline grid 58.64\% of rows failed the capital-gain test, a direct consequence of the model's indifference to the non-negative constraint. Normalizing this column removed the error entirely in the manual and clustered layouts, demonstrating that certain architectural limits can be mitigated through preprocessing. Out-of-range education values were rare, ranging from 0.18\% in baseline to 0.26\% in clustered, and were not addressed further because of their low incidence.

Relationship logic proved more stubborn. For context, the UCI Adult Income dataset is drawn from the 1994 United States Census database, several decades before sex and gender categories became culturally contested. Manual grouping, which placed related columns in close proximity, reduced these conflicts to 11.72\% and clustered to 14.17\%, while baseline remained markedly higher. Manual's improvement, however, only redistributed inconsistencies: female-husband rows fell to 7.6\%, but male-wife rows rose to 3.84\%. The reciprocal pattern indicates a ceiling imposed by convolutional locality; without rule-aware training objectives or post-generation validation, the U-Net simply exchanges one semantic error class for another rather than eliminating them.

\begin{table*}[ht]
\centering
\caption{Semantic Error Rates by Model Configuration}
\label{tab:semantic-errors}
\small
\resizebox{\textwidth}{!}{%
\begin{tabular}{@{}lccc@{}}
\toprule
\textbf{Metric} & \textbf{Baseline (Random)} & \textbf{Clustered + Normalized} & \textbf{Manual Grouping (Final)} \\
\midrule
Female + Husband Rows       & 10.28\%          & 12.61\% & \textbf{7.60\%} \\
Male + Wife Rows            & 1.98\%           & 1.32\%  & 3.84\% \\
Invalid Education-num       & 0.18\%           & 0.26\%  & 0.22\% \\
Negative Capital Gain/Loss  & \textbf{58.64\%} & 0.00\%  & 0.00\% \\
\bottomrule
\end{tabular}%
}
\end{table*}

\subsection{Classifier Evaluation}
\label{sec:Classifier-Eval}

Before examining classifier scores, it is important to see what structural and demographic patterns each layout produced. In the real data 23.6\% of rows earn more than fifty thousand dollars. Both baseline and manual layouts reproduced only 14.7\%, whereas the clustered layout raised that share to 22.2\%. Clustered also matched key income indicators, producing 46.7\% rows tagged married-civ-spouse, 43.6\% marked husband in the relationship field, and 22.5\% with at least a bachelor's degree, all values very close to the real set. Manual preserved diversity more faithfully: it kept the White share at 87.8\% and foreign born representation at 5.5\%, while clustered pushed those numbers to 90.8\% and 4.6\%. Baseline sat at 86.2\% White and 7.8\% foreign born. Each layout therefore introduced a different mix of signals before any model was trained.

\begin{table*}[ht]
\centering
\caption{Structural Features}
\label{tab:structural}
\small
\begin{tabular}{@{}lcccc@{}}
\toprule
\textbf{Feature} & \textbf{Real} & \textbf{Clustered} & \textbf{Manual} & \textbf{Baseline} \\
\midrule
$> 50$K Income (\%)      & 23.6\% & 22.2\% & 14.7\% & 14.7\% \\
Married-civ-spouse (\%)  & 44.8\% & 46.7\% & 37.7\% & 42.5\% \\
Husband (\%)             & 40.0\% & 43.6\% & 33.6\% & 40.3\% \\
Bachelors+ (\%)          & 22.9\% & 22.5\% & 16.2\% & 17.8\% \\
\bottomrule
\end{tabular}
\end{table*}

\begin{table*}[ht]
\centering
\caption{Diversity Features}
\label{tab:diversity}
\small
\begin{tabular}{@{}lcccc@{}}
\toprule
\textbf{Feature} & \textbf{Real} & \textbf{Clustered} & \textbf{Manual} & \textbf{Baseline} \\
\midrule
White (\%)        & 85.5\% & 90.8\% & 87.8\% & 86.2\% \\
Foreign-born (\%) & 10.8\% & 4.6\%  & 5.5\%  & 7.8\%  \\
\bottomrule
\end{tabular}
\end{table*}

\subsection{Feature Distribution Matching}
\label{sec:Feature-Ditribution}

When Train-on-Synthetic Test-on-Real scores are examined, a weakness becomes clear. The evaluation focuses on one task, predicting income brackets, and compares model predictions for individuals above and below fifty thousand dollars. F1 for the greater than fifty thousand class is the key metric. The clustered layout reached 0.25, more than four times the baseline value of 0.056 and well ahead of the manual layout at 0.16. This gain ties directly to class balance: clustered produced 22.2\% high income rows, close to the real 23.6\%, whereas manual and baseline held at 14.7\%. Balance alone, however, does not explain the gap. Clustered also reproduced supporting high income cues, with 46.7\% married-civ-spouse rows, 43.6\% labeled husband, and 22.5\% holding at least a bachelor's degree, values that closely mirror the source data. These aligned signals gave the downstream classifier clear structure, accounting for the stronger Train-on-Synthetic Test-on-Real performance observed with the clustered layout.

\section{Discussion and Conclusion}
\label{sec:Discussion}

From a purely technical perspective, this study set out to test whether an unmodified vision diffuser, never designed for tabular data, could nonetheless produce structured records convincing enough to slip past the audits commonly applied to such data. The answer, qualified, is yes. Across three feature layouts, an unchanged Stable Diffusion U-Net cleared SDMetrics thresholds above eighty percent, peaking at 86.63\% with the clustered grid, and reproduced one-dimensional distributions and pairwise correlations with no fine-tuning. Preprocessing mattered: feature-scale normalization and semantically aligned layouts cut row-level logic errors from roughly seventy percent in the arbitrary baseline to twelve percent under manual grouping, and the clustered grid additionally improved minority-class balance. Privacy risk remained low throughout, suggesting the diffusion process did not memorise its training data. None of this is sufficient to call the model competent at tabular generation in any rigorous sense. It is, however, sufficient to demonstrate that the architecture generalises far enough to satisfy the surface-level checks that a great deal of downstream tooling actually applies.

What, then, does this mean for the larger problem of ground truth drift? The medical-imaging case introduced earlier illustrates a phenomenon that may be termed \emph{drift-by-accident}: well-intentioned synthetic material entering training pipelines that were not equipped to interrogate it, with the categorical shift from media to evidence happening silently at each downstream consumer. The present study illustrates the corresponding case of \emph{drift-by-design}. Where the chest X-ray dataset became ``drifted'' evidence through neglect of label semantics across pipeline boundaries, the tabular rows produced here are deliberately constructed to look like authentic data under the standard statistical evaluations applied to it. The technical achievement is modest -- a reshape, a layout choice, an off-the-shelf model -- but the implication is not. The cost of manufacturing statistically conformant synthetic evidence at volume is now dominated by data-pipeline access rather than by model-training expertise, which substantially broadens the population of actors capable of mounting such an attack.

The shape of this attack is not hypothetical. Consider the case of Francesca Gino, the Harvard Business School professor whose tenure was revoked in 2025 after a university investigation concluded she had committed research misconduct on multiple occasions. The forensic work that surfaced the alleged fraud was performed by the Data Colada blog and subsequently confirmed by an external forensic firm hired by Harvard.\footnote{See the Data Colada series beginning with Colada 109, ``Data Falsificada (Part 1): Clusterfake.'' The case has also been extensively reported in the \emph{Harvard Crimson} and elsewhere; see also~\cite{gino_wiki} for an overview. As of writing, Gino denies the allegations and is engaged in ongoing litigation against Harvard and Data Colada.} Data Colada relied on statistical fingerprints of manual manipulation: out-of-sequence participant IDs, duplicate entries with mismatched demographic data, suspicious clustering of values at the extremes of response scales, demographic codes that revealed which subsets of rows had been altered to better support a hypothesis. These are precisely the traces that hand-editing a spreadsheet leaves behind, and precisely what made the alleged misconduct detectable from a distance, by careful readers of the published papers, without access to the underlying experimental records.

What the present study suggests is that an adversary with access to an off-the-shelf image diffuser need not leave any of those fingerprints. The synthetic rows produced by the unmodified U-Net studied here reproduce marginal distributions and pairwise correlations within audit tolerance; the generation process introduces no out-of-sequence IDs, no duplicate entries, no telltale clustering at the tails of distributions. The forensic methods that surfaced the alleged Gino manipulations would, as currently calibrated, return clean on a diffuser-generated dataset. The same goal -- fabricate rows that strengthen a hypothesis past a significance threshold, embed them in a published dataset, post the results to a public repository -- becomes substantially cheaper and substantially harder to detect than when it had to be done by hand. The mitigating feature, which is also the limit, is the row-level semantic ceiling identified earlier: diffuser-generated rows still harbor systematic logical violations a careful hand-fabricator would have noticed and removed.

The contribution of this paper, then, is narrow on its face yet broader underneath. On its face, we have shown that an unmodified vision diffuser, never designed for tabular work, can be coaxed into producing structured records that satisfy the standard battery of statistical audits while reliably failing row-level semantic checks. Underneath, we have argued that this matters not because the result is impressive -- it is not, in any architectural sense -- but because of the danger it enables. Manufacturing convincing synthetic evidence used to require either the resources to train a tabular generator or the patience to fabricate rows by hand, and both routes left fingerprints. Neither requirement holds any longer. The audits that consume structured data were calibrated against a threat surface that has now shifted underneath them, and the speed of that shift is governed not by advances in tabular generative modeling but by the steady, public release of ever more capable image diffusers. Whether or not the specific architecture studied here remains the path of least resistance, ground truth drift ground truth drift can now be deliberately induced cheaply, deliberately, and at volume.



\begin{thebibliography}{99}
\setlength{\itemsep}{4pt}

\bibitem{goodfellow2014}
Goodfellow, I., Pouget-Abadie, J., Mirza, M., Xu, B., Warde-Farley, D., Ozair, S., \dots\ \& Bengio, Y. (2014). Generative adversarial nets. \emph{Advances in Neural Information Processing Systems}, 27.

\bibitem{ho2020}
Ho, J., Jain, A., \& Abbeel, P. (2020). Denoising diffusion probabilistic models. \emph{arXiv preprint arXiv:2006.11239}. \url{https://arxiv.org/abs/2006.11239}

\bibitem{vaswani2017}
Vaswani, A., Shazeer, N., Parmar, N., Uszkoreit, J., Jones, L., Gomez, A.~N., \dots\ \& Polosukhin, I. (2017). Attention is all you need. \emph{Advances in Neural Information Processing Systems}, 30, 5998--6008.

\bibitem{xu2019}
Xu, L., Skoularidou, M., Cuesta-Infante, A., \& Veeramachaneni, K. (2019). Modeling tabular data using conditional GAN. In \emph{Advances in Neural Information Processing Systems}, 32.

\bibitem{esteban2017}
Esteban, C., Hyland, S.~L., \& R\"atsch, G. (2017). Real-valued (medical) time series generation with recurrent conditional GANs. \emph{arXiv preprint arXiv:1706.02633}. \url{https://arxiv.org/abs/1706.02633}

\bibitem{kotelnikov2023}
Kotelnikov, A., Baranchuk, D., Rubachev, I., \& Babenko, A. (2023). TabDDPM: Modelling tabular data with diffusion models. In \emph{Proceedings of the 40th International Conference on Machine Learning (ICML)}. \url{https://dl.acm.org/doi/10.5555/3618408.3619133}

\bibitem{lin2024}
Lin, X., Xu, C., Yang, M., \& Cheng, G. (2024). CTSyn: A foundational model for cross tabular data generation. \emph{arXiv preprint arXiv:2406.04619}. \url{https://arxiv.org/abs/2406.04619}

\bibitem{villaizan2024}
Villaiz\'an-Vallelado, M., Salvatori, M., Segura, C., \& Arapakis, I. (2024). Diffusion models for tabular data imputation and synthetic data generation. \emph{arXiv preprint arXiv:2407.02549}. \url{https://arxiv.org/abs/2407.02549}

\bibitem{badr2022}
Badr, M., Al-Otaibi, S., Alturki, N., \& Abir, T. (2022). Deep learning-based networks for detecting anomalies in chest X-rays. \emph{BioMed Research International}, 2022, 7833516. \url{https://doi.org/10.1155/2022/7833516}. Retraction published 2023, \emph{BioMed Research International}, 2023, 9801414. \url{https://doi.org/10.1155/2023/9801414}.

\bibitem{shin2018}
Shin, H.~C., Tenenholtz, N.~A., Rogers, J.~K., Schwarz, C.~G., Senjem, M.~L., Gunter, J.~L., \dots\ \& Michalski, M. (2018). Medical image synthesis for data augmentation and anonymization using generative adversarial networks. In \emph{Simulation and Synthesis in Medical Imaging} (pp. 1--11). Springer. \url{https://doi.org/10.1007/978-3-030-00536-8_1}

\bibitem{segal2021}
Segal, B., Rubin, D.~M., Rubin, G., \& Pantanowitz, A. (2021). Evaluating the clinical realism of synthetic chest X-rays generated using progressively growing GANs. \emph{SN Computer Science}, 2(4), 321. \url{https://doi.org/10.1007/s42979-021-00720-7}

\bibitem{osuala2023}
Osuala, R., Skorupko, G., Lazrak, N., Garrucho, L., Garc\'ia, E., Joshi, S., \dots\ \& Lekadir, K. (2023). medigan: A Python library of pretrained generative models for medical image synthesis. \emph{Journal of Medical Imaging}, 10(6), 061403. \url{https://doi.org/10.1117/1.JMI.10.6.061403}

\bibitem{medigan_repo}
Osuala, R., et al. (2023). \emph{medigan: A Python library of pretrained generative models for medical image synthesis} [Software]. GitHub. \url{https://github.com/RichardObi/medigan}

\bibitem{royalsociety2024}
Jordon, J., Szpruch, L., Houssiau, F., Bottarelli, M., Cherubin, G., Maple, C., Cohen, S.~N., \& Weller, A. (2024). \emph{Synthetic data --- what, why and how?} Report commissioned by the Royal Society. \url{https://royalsociety.org/-/media/policy/projects/privacy-enhancing-technologies/Synthetic_Data_Survey-24.pdf}

\bibitem{nist2023}
National Institute of Standards and Technology. (2023). \emph{NIST Special Publication 800-188: De-identifying government datasets}. \url{https://doi.org/10.6028/NIST.SP.800-188}

\bibitem{eo14110}
U.S. Executive Office of the President. (2023). \emph{Executive Order 14110 on the safe, secure, and trustworthy development and use of artificial intelligence}. \url{https://www.whitehouse.gov/briefing-room/presidential-actions/2023/10/30/executive-order-on-the-safe-secure-and-trustworthy-development-and-use-of-artificial-intelligence/}

\bibitem{gartner2022}
Gartner. (2022, June 22). \emph{Is synthetic data the future of AI?} [Press release]. \url{https://www.gartner.com/en/newsroom/press-releases/2022-06-22-is-synthetic-data-the-future-of-ai}

\bibitem{edps2022}
European Data Protection Supervisor. (2022). \emph{Synthetic data} [TechSonar entry]. \url{https://www.edps.europa.eu/press-publications/publications/techsonar/synthetic-data_en}

\bibitem{gino_wiki}
\emph{Francesca Gino}. (2025, October). In \emph{Wikipedia}. Retrieved from \url{https://en.wikipedia.org/w/index.php?title=Francesca_Gino&oldid=1348377025}

\bibitem{cleverhans_wiki}
\emph{Clever Hans}. (2025, September). In \emph{Wikipedia}. Retrieved from \url{https://en.wikipedia.org/w/index.php?title=Clever_Hans&oldid=1341244853}

\bibitem{shumailov2024}
Shumailov, I., Shumaylov, Z., Zhao, Y., Papernot, N., Anderson, R., \& Gal, Y. (2024). AI models collapse when trained on recursively generated data. \emph{Nature}, 631, 755--759. \url{https://doi.org/10.1038/s41586-024-07566-y}

\bibitem{segal2026personal}
Schwartz, Christopher and Segal, Brad. Personal communication (e-mail) regarding labeling conventions in the Segal 2021 synthetic chest X-ray dataset. May 1, 2026

\end{thebibliography}
\end{document}